\documentstyle[aaspp4,flushrt]{article}

\begin{document}

\title{The Mass distribution of the Most Luminous X-ray Cluster RXJ1347.5-1145
from Gravitational Lensing}

\author{Philippe Fischer\footnote {Visiting Astronomer, National Optical
Astronomy Observatories, which is operated by the Association of Universities
for Research in Astronomy, Inc., under contract to the National Science
Foundation.}\footnote{Hubble Fellow} }

\affil{Dept. of Astronomy, University of Michigan, Ann Arbor, MI 48109}

\author{J. Anthony Tyson} 

\affil{Bell Laboratories-Lucent Technologies, 700 Mountain Ave., Murray Hill,
NJ 07974}

\abstract

Galaxy cluster mass distribution are potentially useful probes of $\Omega_0$
and the nature of the dark matter. Large clusters will distort the observed
shapes of background galaxies through gravitational lensing allowing the
measurement of the cluster mass distributions. In this paper we describe weak
statistical lensing measurements of the most luminous X-ray cluster known,
RXJ1347.5-1145 at $z=0.45$. We detect a shear signal in the background galaxies
at a signal-to-noise ratio of 7.5 in the radial range $120 \le r \le 1360
h^{-1}$ kpc. A mass map of the cluster reveals an 11$\sigma$ peak in the
cluster mass distribution consistent with the position of the central dominant
galaxy and 3 $\sigma$ evidence for the presence of a subcluster at a projected
radius of 1.3 - 1.7 $h^{-1}$ Mpc from the cluster center. In the range $120 \le
r \le 1360 h^{-1}$ kpc mass traces light, and the azimuthally averaged cluster
mass and light profiles are consistent with singular isothermal spheres with
M$(r<1$ Mpc) = $1.7 \pm 0.4 \times 10^{15}$ M$_\odot$. Assuming an isotropic
velocity distribution function, the implied velocity dispersion is $\sigma =
1500 \pm 160$ km s$^{-1}$. The rest-frame mass-to-light ratio is M/L$_B = 200
\pm 50 h$ M$_\odot$/L$_{B\odot}$. 
The lensing mass estimate is almost twice as high as a previously determined
X-ray mass estimate.

\section{Introduction}






Galaxy clusters are useful probes of large scale structure, and their mass
distributions reflect the underlying cosmology (Richstone et
al. 1992). Specific predictions are emerging from large scale structure
simulations regarding the mass distributions in clusters, the degree of
subclustering, and the relationship between light and mass and how these vary
when cosmological parameters ($\Omega$, slope of the density perturbation
spectrum, type of dark matter particles, etc) are varied. Accurate surface mass
density measurements for a sample of clusters can thus place strong constraints
on $\Omega$ and the nature of the dark matter (Crone et al. 1994, Crone et
al. 1996, Wilson et al 1996). Cluster mass measurements are also useful tests
of big bang nucleosynthesis; coupled with gas and stellar mass estimates, the
ratio of baryonic matter to dark matter, $\Omega_B$, can be estimated (White
\& Fabian 1995, Evrard 1997).

The relative numbers of massive clusters found at redshifts $z=0.2$ and $z=0.5$
is a sensitive test of $\Omega$.  In this redshift range rich clusters evolve
in mass much more rapidly in high $\Omega$ universes; at $z = 0.5$ seven times
as many massive clusters are expected for $\Omega = 0.3$ than for $\Omega = 1$,
while at $z = 0.2$ about four times as many massive clusters are expected for
$\Omega = 1$ than for $\Omega = 0.3$, independent of $\Lambda$ (Eke et al
1996). So it is important to measure the masses of clusters at $z = 0.5$ as
well as $z = 0.2$.

The study of the mass distributions in galaxy clusters via gravitational
lensing is now a well established technique (Tyson et al 1990, Tyson \& Fischer
1995, Squires et. al. 1996a, 1996b, 1996c and Fischer et al. 1996). Lensing
provides the only direct means of measuring cluster masses; unlike dynamical
and X-ray mass estimates it does not depend on knowledge of the dynamical state
of the matter.

The cluster RXJ 1347.5-1145, at a redshift of $z=0.451$, is the most luminous
X-ray cluster known with a bolometric luminosity L$_X = 2 \times 10^{46}$ erg
s$^{-1}$ (Schindler et al. 1995).
As reported by Schindler et al. (1995) there are two arc candidates at
projected angular distances of around 35\arcsec\ (120 h$^{-1}$ kpc) from the
central dominant galaxy (CDG), which implies a large amount of mass contained
within that radius. However, the X-ray mass estimate (Schindler et al. 1996) is
much lower than implied by the arcs, and is also lower than would be naively
expected from the total X-ray luminosity.

We have obtained deep optical CCD images of RXJ1347.5-1145 in two bandpasses
over a wide field in order to study the cluster mass distribution by its
distorting effect on the shapes of background galaxies. These observations are
described in \S \ref{observations}.  In \S \ref{profit} we discuss the galaxy
photometry and shape analysis. \S \ref{anisotropy} describes the technique used
to correct for PSF anisotropy. \S \ref{shear} shows the radial shear profile of
the cluster and \S \ref{mass} describes the equations for one and
two-dimensional mass reconstruction. \S \ref{calibration} contains a
description of simulations carried out to quantify the dilution due to seeing
and \S \ref{mass} is a discussion of the cluster mass distribution, total mass
and mass-to-light ratio. \S \ref{mass} also has a comparison of the X-ray mass
estimate and the lensing mass estimate.

\section{Observations} \label{observations}

The cluster RXJ1347-1145 was observed using the prime focus CCD camera on the
Blanco 4m at CTIO on 26 June 1995. The total exposure time was 4800s in B$_J$
($16 \times 300$s) and 3300s in R ($11 \times 300$s). The telescope was
dithered between exposures.  The ``Tek4'' 2048$^2$ thinned SITE CCD was used
with 0.43\arcsec\ pixels. The seeing on the combined B$_J$ image averaged
1.2\arcsec\ FWHM and 1.25\arcsec\ for the combined R image, but was variable
across the images (see \S \ref{anisotropy}). The night was not photometric so
the images were normalized prior to combining.  Additional images were obtained
on the two previous nights, however, due to poor image quality ($\approx
2\arcsec$ FWHM), they were not used. The total useable field size in the
combined images is approximately $14.3\arcmin\ \times 14.0\arcmin$ for B$_J$
and $14.5\arcmin \times 14.2\arcmin$ for R. The RMS sky noise is 29.1 B$_J$ mag
per square arcsec and 27.7 R mag per square arcsec. A color image of a
subsection of the the CCD field centered on the cluster is shown in
Fig. \ref{color}.

\begin{figure}
\caption{Color image of the RXJ1347.5-1145 field
constructed from B$_J$ and R band images. This extract of our larger CCD field
measures 6.8 arcminutes on a side. The log of the intensity is shown.
North is up and East is to the left.
\label{color}}
\end{figure}

\section{Faint Galaxy Photometry and Analysis} \label{profit}

The faint galaxy analysis was carried out using the analysis software ProFit
(developed by PF). This software, starting with the brightest objects, fits an
analytical model to each, using weighted, non-linear least squares, and
subtracts the light from the image. It then proceeds to successively fainter
objects. Once it has detected and subtracted all the objects in an image it
replaces each in turn and refits and resubtracts until convergence is
achieved. The software outputs brightness, orientation, ellipticity and other
image parameters based on the fitted function.

\begin{figure}
\plotone{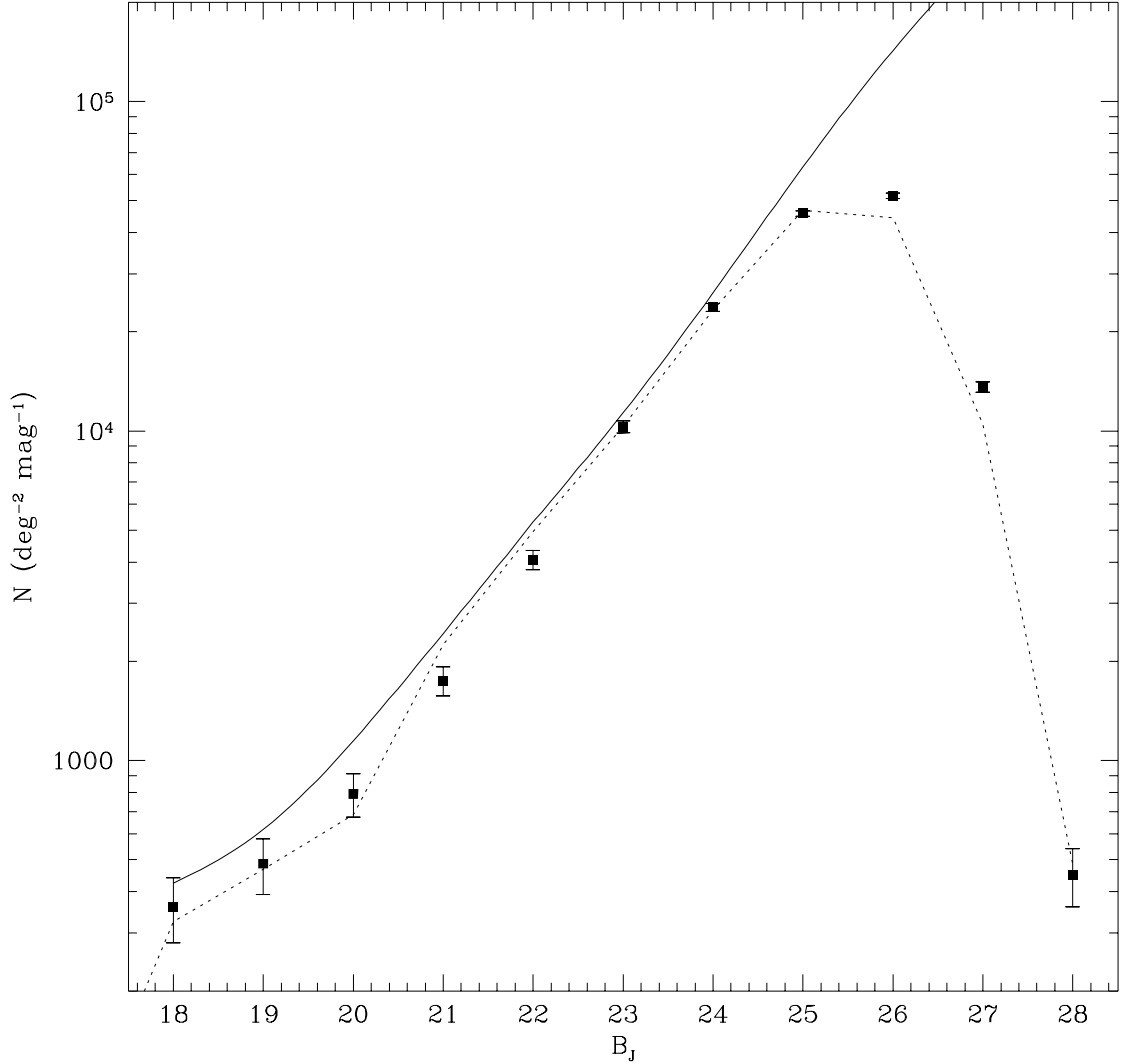}
\caption{Counts for objects (stars and galaxies) detected
in the field of RXJ 1347.5-1145. The points are the data, the solid line is the
input model for the simulations and the dashed line is the result from one of
the simulations. The magnitudes are the total magnitude within an outer
isophote of 28.9 B$_J$ mag per square arcsecond
\label{lumfunc}}
\end{figure}

Fig. \ref{lumfunc} shows B$_J$ galaxy counts for the faint galaxies in the
field of RXJ 1347-1145. The magnitudes are isophotal magnitudes with outer
isophote of 28.9 B$_J$ mag square arcsecond (30.8 B$_J$ mag per pixel).

The lower panel of Fig. \ref{cmd} shows a B$_J$--R color magnitude diagram of
the whole field. There are a large number of red galaxies with an indication of
a gap around B$_J-$ R = 1.6 mag. The upper panel shows only the inner
90\arcsec\ and one can now clearly see the cluster early-type galaxy sequence
centered around B$_J - $R = 2.3 mag.

\begin{figure}
\caption{B$_J$--R color-magnitude diagram of objects in the
field around RXJ1347.5-1145. The B$_J$ magnitudes are isophotal magnitudes (see
Fig. \protect\ref{lumfunc}) and the colors are derived in an aperture of
1.3\arcsec\ radius. The lower panel is the whole field containing 5870 objects
while the lower panel is the inner 90\arcsec\ (303 objects).
\label{cmd}}
\end{figure}

\section{PSF Anisotropy} \label{anisotropy}

\begin{figure}
\plotone{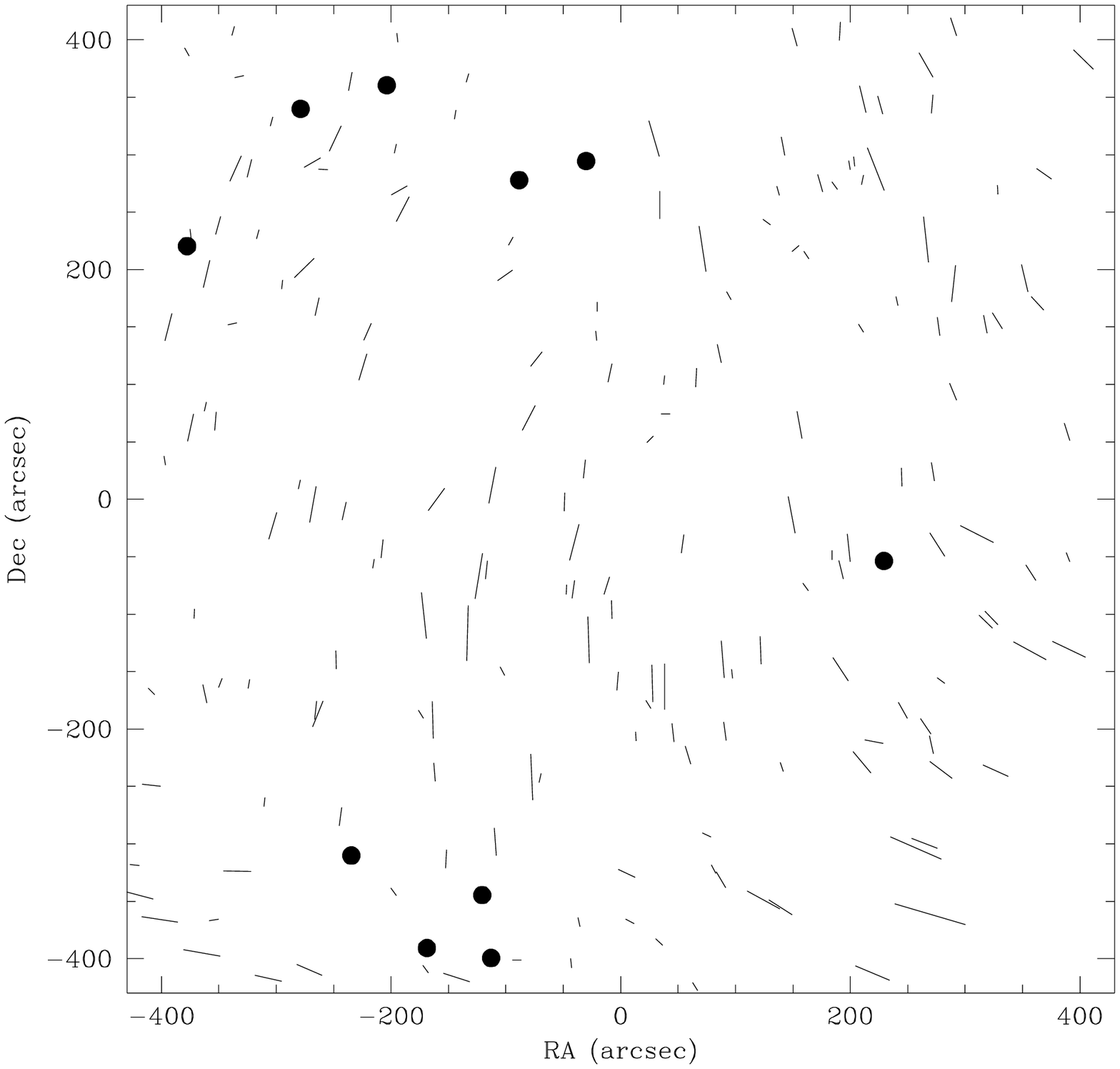}
\caption{Ellipticity and orientation for 193 stars in the
combined B$_J$-band image. Solid circles indicate stars which have $\epsilon <
0.005$. The maximum ellipticity is $\epsilon = 0.08$ and the mean is
$<\epsilon> = 0.02$. The shapes and orientations of the stars are clearly
correlated with position in this image. North is up and East is to the left.
\label{psf1}}
\end{figure}

There are several factors which can contribute to PSF anisotropy (Kaiser et al
1995). In June 1995 at the CTIO 4m there were at least two known contributors,
telescope astigmatism (and possible misalignment of the prime focus corrector)
and a 200 micron warp in the Tek 2k CCD. The latter means that it is impossible
to focus the whole CCD, and because of the former, elliptical PSFs
result. Fig. \ref{psf1} shows the ellipticity and orientation for 193 stars in
the combined B$_J$-band image of RXJ1347-1145. As expected the shape and
orientation of the stars are correlated with position on the image. Since the
distortion due to weak lensing is very small, this PSF anisotropy, if left
uncorrected, will complicate the interpretation of the data.

For this paper we choose to correct for PSF anisotropy by devising a
position-dependent convolution kernel which circularizes the PSF.  After
convolving the image with the kernel the PSFs should be substantially rounder
and large scale correlations of shape and orientation will be removed.
Unfortunately, a circularizing kernel will also blur the PSF somewhat, so we
constrain the kernel to minimize this blurring effect. Additional constraints
are that it be flux conserving and have all elements greater than or equal to
zero.

The first step in constructing the kernel is to measure the usual quadratic
moments for all the stars:

\begin{eqnarray}
I_{xx}(PSF) & = & {\sum{F_ix_i^2} \over \sum{F_i}}, \nonumber \\
I_{yy}(PSF) & = & {\sum{F_iy_i^2} \over \sum{F_i}}, \\
I_{xy}(PSF) & = & {\sum{F_ix_iy_i} \over \sum{F_i}}, \nonumber
\end{eqnarray}

\noindent where $F_i$ is the intensity of pixel $i$ minus the sky background
intensity. A kernel ($K$) which satisfies our criteria and will circularize a
PSF having $I_{xx}(PSF), I_{yy}(PSF),$ and $I_{xy}(PSF)$ has:

\begin{eqnarray}
\left.
\begin{array}{ll}
I_{xx}(K) =  & |I_{xy}(PSF)| \\
I_{yy}(K) =  & I_{xx}(PSF) - I_{yy}(PSF) + |I_{xy}(PSF)| \\
I_{xy}(K) =  & -I_{xy}(PSF) \nonumber \\
\end{array}
\right\} I_{xx}(PSF) > I_{yy}(PSF) \nonumber \\
\nonumber \\ 
\nonumber \\ 
\left.
\begin{array}{ll} \nonumber
I_{xx}(K) =  & I_{yy}(PSF) - I_{xx}(PSF) + |I_{xy}(PSF)| \nonumber\\
I_{yy}(K) =  & |I_{xy}(PSF)| \nonumber \\
I_{xy}(K) =  & -I_{xy}(PSF) \nonumber \\
\end{array} \nonumber
\right\} I_{xx}(PSF) < I_{yy}(PSF) \nonumber \\
\end{eqnarray}


\noindent 
The goal is to produce a kernel which, when convolved with the original image,
will yield PSFs with $I_{xx}=I_{yy}$ and $I_{xy} = 0$, within measurement
error. The corresponding nine element convolution kernel which minimizes image
blurring, is flux conserving, and positive everywhere has elements:

\begin{eqnarray} \label{kerneqn}
K(1,1) & = & K(3,3) = 0.25[I_{xx}(K)+I_{yy}(K)+I_{xy}(K)-1+K(2,2)] \nonumber\\
K(1,2) & = & K(3,2) = 0.5[1-I_{yy}(K)-K(2,2)] \nonumber\\
K(2,2) & = & min[1-I_{xx}(K),1-I_{yy}(K)]\\
K(3,1) & = & K(1,3) = 0.25[I_{xx}(K)+I_{yy}(K)-I_{xy}(K)-1+K(2,2)] \nonumber\\
K(2,3) & = & K(2,1) = 0.5[1-I_{xx}(K)-K(2,2)] \nonumber
\end{eqnarray}

\begin{figure}
\plotone{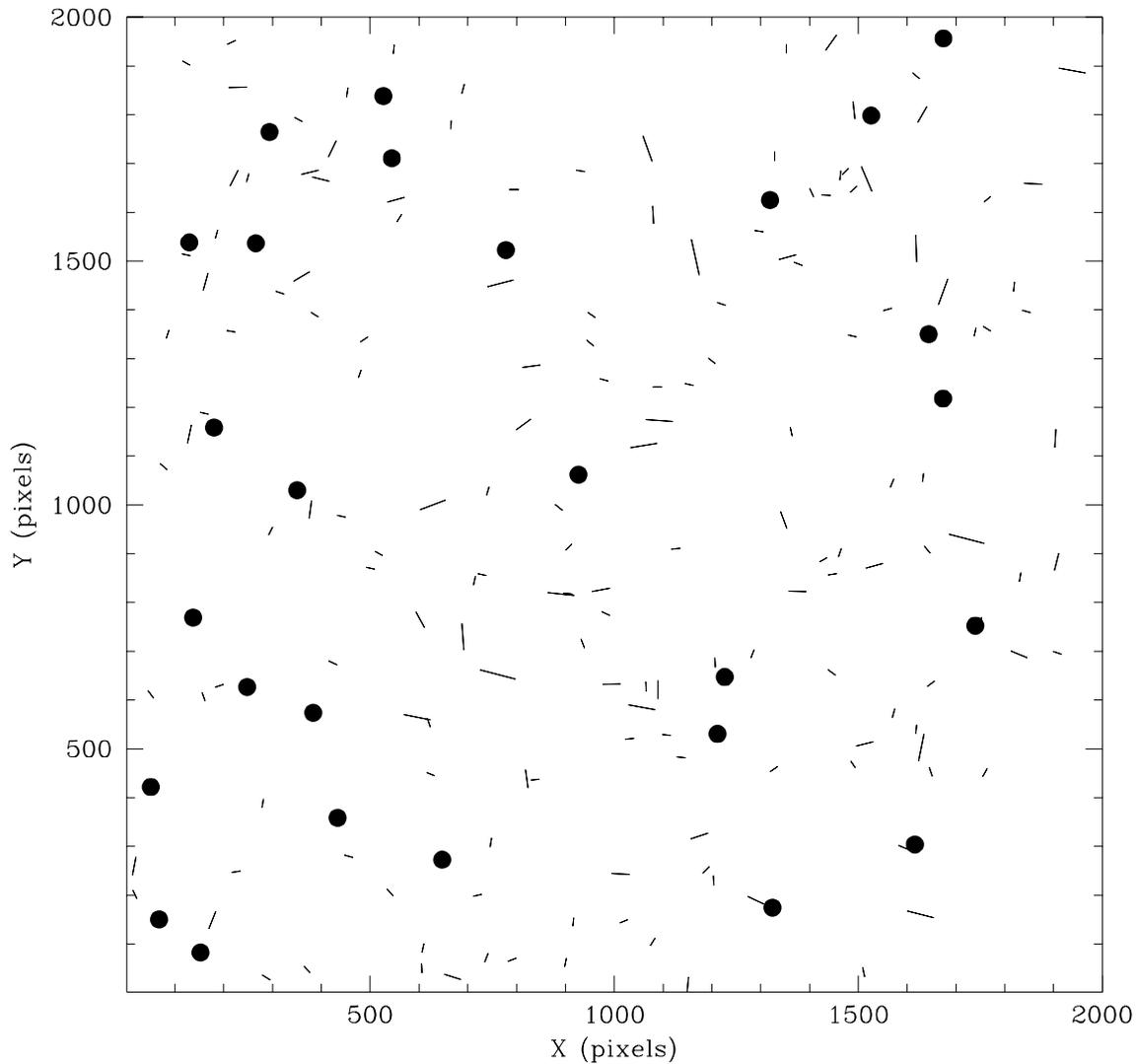}
\caption{Ellipticity and orientation for the same 193 stars
as in Fig. \protect\ref{psf1} after convolution with a position dependent
kernel designed to circularize the PSF. Solid circles indicate stars which have
$\epsilon < 0.005$. The maximum ellipticity is $\epsilon = 0.04$ and the mean
is $<\epsilon> = 0.01$. The shape and orientation are no longer correlated with
position. \label{psf2}}
\end{figure}

As can be seen from \ref{kerneqn} the elements of the kernel are only dependent
on $I_{xy}(PSF)$ and $I_{xx}(PSF) - I_{yy}(PSF)$.  In order to correct the CCD
image we fit third order polynomials to $f_1(x,y)=[I_{xx}(PSF)-I_{yy}(PSF)]$
and $f_2(x,y)=I_{xy}(PSF)$. If we know the values of $f_1$ and $f_2$ everywhere
on the image we can construct a position dependent kernel using the
prescription given by \ref{kerneqn}.  We then convolve our image with the
kernel removing most of the PSF anisotropy. Fig. \ref{psf2} shows the PSFs
after the convolution; the mean measured ellipticity has been reduced from
$<\epsilon> = 0.02$ to $<\epsilon> = 0.01$ and the shapes and orientation are
no longer correlated with position. This is quantified further in \S
\ref{shear} and \S \ref{2d}.

\section{Shear} \label{shear}

For gravitational lensing, the relationship between the tangential shear,
$\gamma_T$, and surface mass density, $\Sigma$, is (Miralda-Escud\'e 1991,
1995),

\begin{equation}\label{escude}
\gamma_T(r) = \overline{\kappa}(\le r) - \overline{\kappa}(r),
\end{equation}

\noindent
where $\kappa = \Sigma/\Sigma_{crit}$, the ratio of the surface density to the
critical surface density for multiple lensing, and $r$ is the angular distance
from a given point in the mass distribution. The critical density depends on
the redshift distribution of the background galaxies. The first term on the
right is the mean density interior to $r$ and the second term is the mean
density at $r$. Therefore, the presence of a foreground mass distribution will
distort the appearance of background galaxies.  For a given coordinate on the
image ($\vec{r}$), the distortion quantity for the i$^{th}$ galaxy is:

\begin{equation} 
D_i(\vec{r}) = {1-(b_i/a_i)^2 \over 1 + (b_i/a_i)^2} \times
{[\cos(2\theta_i)(\Delta{x_i}^2 - \Delta{y_i}^2) +
2\sin(2\theta_i)\Delta{x_i}\Delta{y_i}] \over \Delta{x_i}^2 + \Delta{y_i}^2},
\end{equation}

\noindent where $(b_i/a_i)$ and $\theta_i$ are the galaxy axis ratio and
position angle, respectively. $\Delta x$ and $\Delta y$ are the horizontal and
vertical angular distances from $\vec{r}$ to galaxy $i$. The value of $D_i$ is
a maximum when the position angle of the $i^{th}$ galaxy major axis is
perpendicular to the line joining $\vec{r}$ to the galaxy and a minimum when it
is parallel. $D$ is related to the tangential shear by (Schneider \& Seitz
1995):

\begin{equation}
<D(r)> = 2{\gamma_T(r)[1-\kappa(r)] \over [1-\kappa(r)]^2 + \gamma_T^2(r)}
\end{equation}

In the weak lensing regime $\kappa << 1$, and $\gamma_T << 1$, $\gamma_T
\approx <D>/2$.

The value of the mean distortion, $<D(\vec{r})>$, for $\vec{r}$ equal to the
position of the CDG in RXJ1347.5-1145 for 1970 galaxies in the combined B$_J$
image having $23.0 \le$ B$_J \le 25.0$ in the radial range $35\arcsec\ \le r
\le 400\arcsec$ is $<D> = 0.031 \pm 0.004$; the signal-to-noise is around 7.5.
For comparison, the value of $<D(\vec{r})>$ for $\vec{r}$ at the CDG location
for the 193 PSF stars in the corrected image is $0.0017 \pm 0.0008$, less than
6\% of the measured shear value.  The PSF-induced distortion on the galaxies is
actually smaller than this since they are resolved objects.  Therefore, the
residual PSF anisotropy has little effect on our cluster mass estimate.


The measured value of $<D>$ is affected by seeing and the shear polarizability
of the galaxies (see Kaiser et al. 1995). In order to calibrate this effect we
carry out simulations using the F450W Hubble Deep Field Data (HDF) (Williams et
al. 1996), using the techniques described in Kaiser et al. (1995). This
involves stretching the HDF data by $1+\delta$, convolving with the PSF and
adding noise. The values of $D_i$ are measured for each galaxy and compared to
the unstretched values of $D_i$. The quantity of interest is the recovery
factor, $C = \delta/<\Delta{D_i}>$. Unfortunately the transformation between
F450W and B$_J$ is dependent on galaxy type and redshift. In order to match the
two bandpasses as closely as possible and therefore sample similar galaxy
populations we choose a 2.0 magnitude range in the HDF data which yields the
same surface number density of galaxies as in the RXJ 1347-1145 field for the
range $23.0 \le$ B$_J \le 25.0$. The resulting F450W magnitude range is $22.5
\le F450W_{HDF} \le 24.5$.  The value of the recovery factor is $<C> = 3.0 \pm
0.15$.  This correction does not take into account the presence of stars which
will further dilute the signal, however, this will be minor for the magnitude
range considered. The diluting effect of cluster galaxies is discussed in \S
\ref{profile}.

\begin{figure}
\plotone{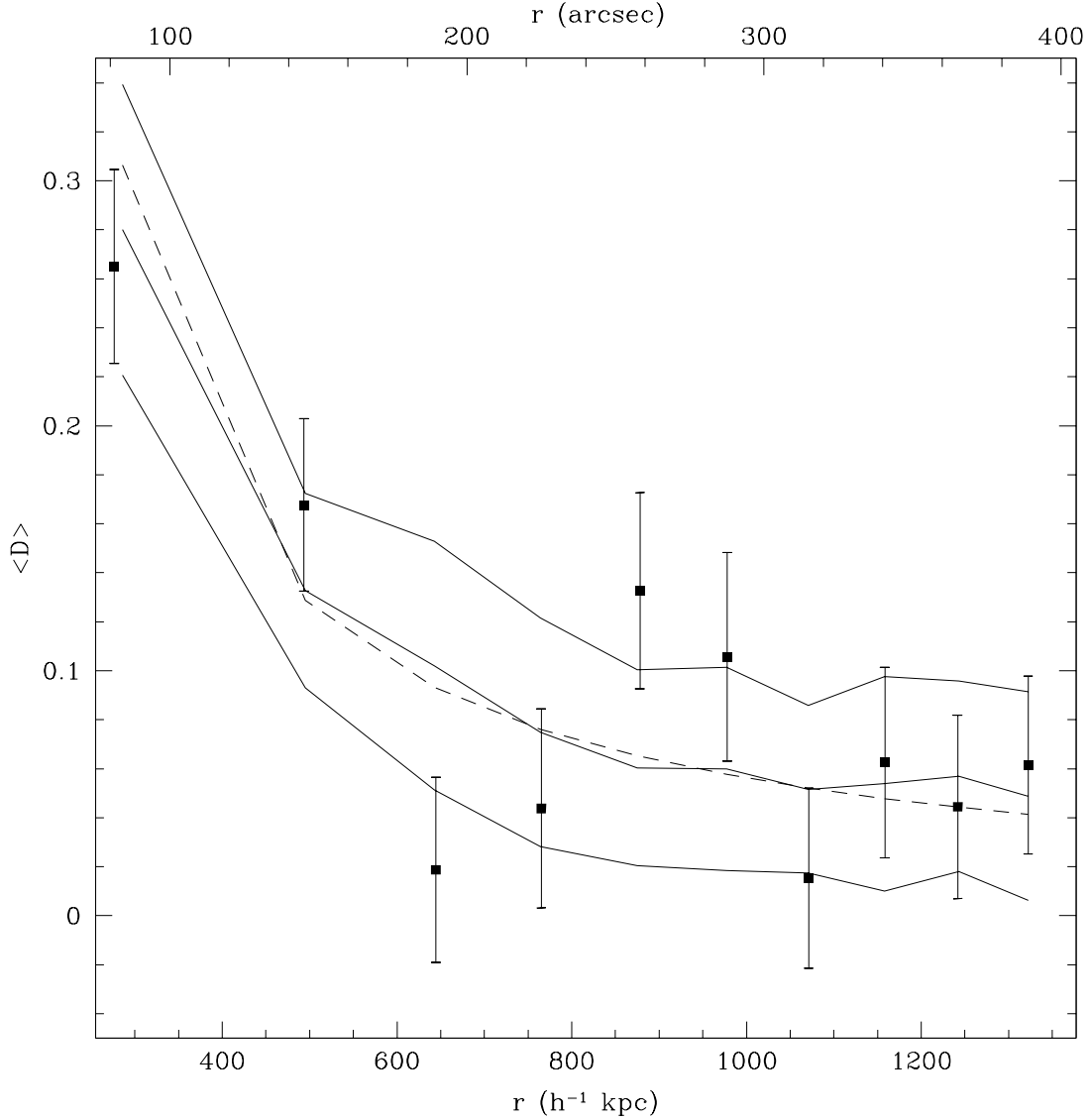}
\caption{A plot of the distortion, $<D>$, vs projected
radius in radial bins containing 197 galaxies each ($35\arcsec\ \le r \le
400\arcsec$, $23.0 \le B_J \le 25.0$), centered on the dominant cluster
galaxy. The points are from the data ($1\sigma$ error bars) and the solid lines
are the mean and $1\sigma$ upper and lower uncertainty bands from 100
Monte-Carlo simulations of a singular isothermal lens with $\sigma=1460$ km
s$^{-1}$. These have been scaled by a recovery factor of $C=3.0$ (see
text). The dashed line is the input model averaged over the radial bins and
corrected for varying $\Sigma_{crit}$.
\label{shearf}}
\end{figure}

Fig. \ref{shearf} shows corrected $<D(\vec{r})>$ vs. projected radius for
$\vec{r}$ at the CDG position. Also shown are the input and measured values for
100 simulations of a lens with a singular isothermal mass distribution with
velocity dispersion $\sigma=1460$ km s$^{-1}$ (see \S
\ref{calibration}). Because the cluster magnifies the background galaxies as

\begin{figure}
\plotone{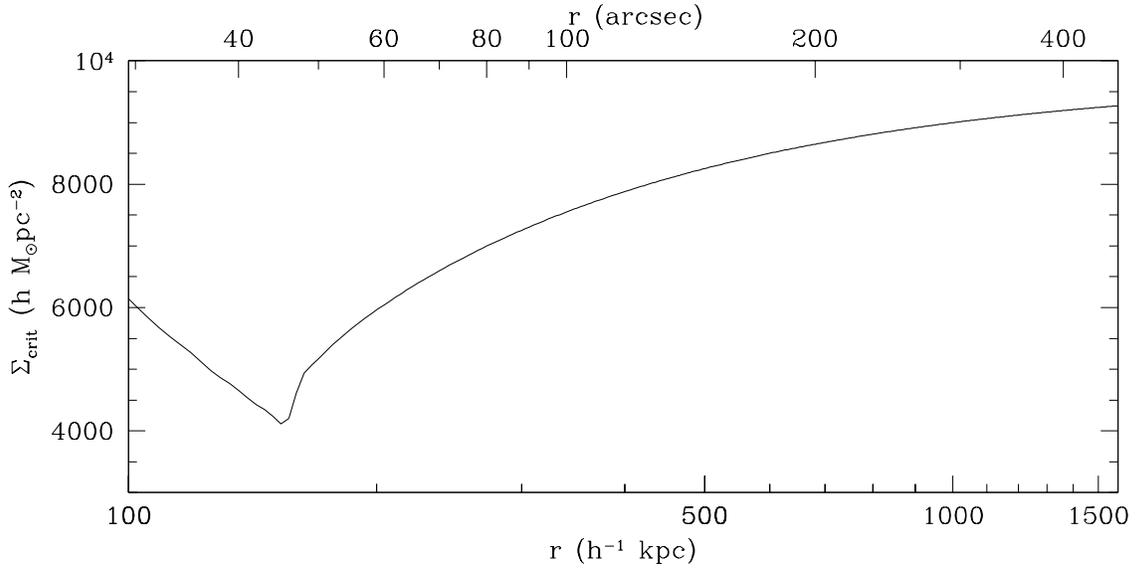}
\caption{$\Sigma_{crit}$ as a function of projected radius
for an isothermal lens having $\sigma = 1460$ km s$^{-1}$, $z=0.45$, and source
galaxies with $23.0 \le $B$_J \le 25.0$. $\Sigma_{crit}$ is dependent on radius
because magnification by the cluster biases the source galaxy redshift upwards.
\label{sigc}}
\end{figure}

\begin{equation}
Mag(r)={1 \over [1-\kappa(r)]^2-\gamma(r)^2}{\ \ ,}
\end{equation}

\noindent the apparent magnitude range which the background galaxies would have
had in the absence of lensing is a function of projected radius from the
cluster center (assuming a circular mass distribution). Since mean redshift
depends on unlensed magnitude, the mean redshift and hence $\Sigma_{crit}$ are
also functions of projected radius. Fig. \ref{sigc} shows how $\Sigma_{crit}$
varies as a function of projected radius for an isothermal lens with $\sigma =
1460$ km s$^{-1}$, $z=0.45$, and source galaxies with $23.0 \le $B$_J \le
25.0$, assuming the galaxy redshift distribution described in \S
\ref{calibration}. The values for $<D>$ for the isothermal model shown in
Fig. \ref{shearf} have been calculated assuming $\Sigma_{crit}$ shown in
Fig. \ref{sigc} (the model used fits the lensing data well, see \S
\ref{profile}). Excellent agreement is seen between the input and measured
$<D>$ values for the simulations, although the measured value is slightly below
the input value for the innermost bin. The innermost bin is the most
susceptible to systematic errors due to the strong lensing occuring in this bin
and the large magnification factor.

The best value for the recovery factor from the simulations is $C = 2.9 \pm
0.05$, in agreement with the HDF simulations (although this agreement is
expected since the simulated galaxy sizes were chosen such that they would
yield the same recovery factor as the HDF simulations). The reduced $\chi^2$
between the singular isothermal model and the data is $\chi^2_{\nu} = 1.4$
($\nu = 9$). Because of the large uncertainties in the binned shear
measurements it is not possible to put strong constraints on the shape of the
mass profile.


\section{Cluster Mass} \label{mass}

\subsection{Mass reconstruction} \label{reconstruct}

Formulae for 2-d mass reconstruction in the weak lensing regime have been
discussed in Kaiser \& Squires (1993) (KS) and Fahlman et al. (1994). Briefly,
in the weak lensing regime, where $\kappa << 1$, the formula for the surface
mass density is:

\begin{equation} \label{kseqn}
\kappa(\vec{r}) = {1\over
\overline{n}\pi}\sum_{i=1}^{N}{W(\Delta{x},\Delta{y},s)
D_i(\vec{r})\over \Delta{x_i}^2 + \Delta{y_i}^2}.
\end{equation}

\noindent 
where $N$ is the number of galaxies and $\overline{n}$ is the number density of
galaxies. Eqn. \ref{kseqn} assumes that the galaxies are intrinsically (in the
absence of lensing) randomly aligned. $W$ is a smoothing kernel which is
required to prevent infinite formal error.  In this paper we use a smoothing
kernel of the form (Seitz \& Schneider 1995):

\begin{equation}
W(\Delta{x},\Delta{y},s)=1-\left(1+{\Delta{x}^2+\Delta{y}^2 \over
2s^2}\right)e^{-(\Delta{x}^2+\Delta{y}^2)/2s^2},
\end{equation}

\noindent
where `$s$' is referred to as the ``smoothing scale''. A 2-d mass map of RXJ
1347.5-1145 is shown in Fig. \ref{massmap} and is discussed further in \S
\ref{2d}.

\begin{figure}
\caption{Mass map derived using Eqn. \protect\ref{kseqn}
with smoothing scale $s = 43\arcsec$. A total of 2735 galaxies with $23.0 \le$
B$_J \le 25.0$ are used in this reconstruction. The contours are spaced in
1$\sigma$ intervals. The peak of the mass distribution is consistent with the
position of the central dominant galaxy. North is up and East is to the
left. The field is 14\arcmin\ on a side.
\label{massmap}}
\end{figure}

\begin{figure}
\plotone{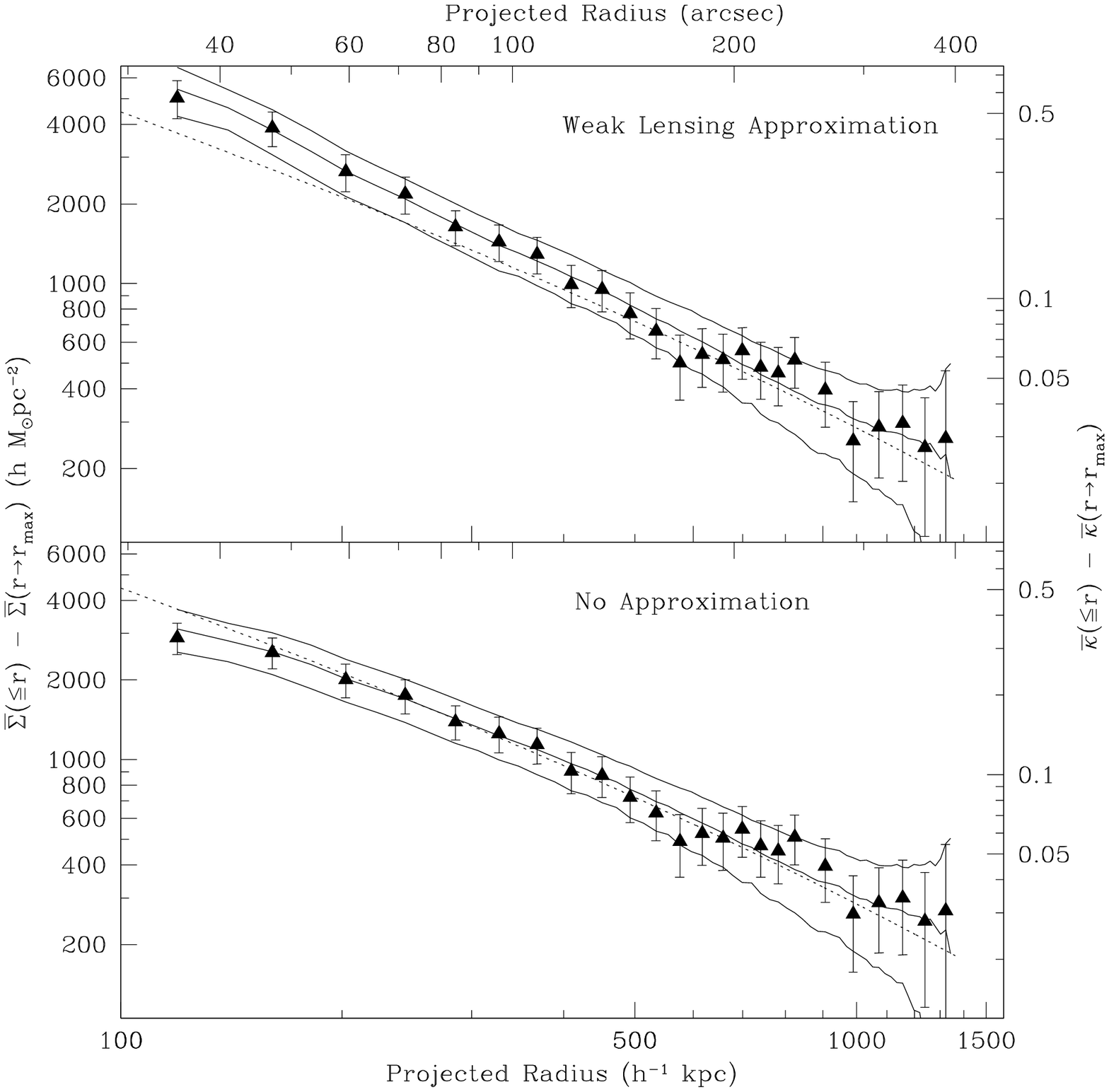}
\caption{The upper panel is the radial mass density profile
for RXJ 1347.4-1145 from Eqn. \protect\ref{denseqn} assuming $\kappa = 0$ and
$\Sigma_{crit}$ is a constant function of radius ($r_{max} = 400\arcsec\ = 1360
h^{-1}$ kpc). It is centered on the central dominant galaxy. The points are the
data for the cluster derived from 1970 galaxies having $23.0 \le$ B$_J \le
25.0$.  The solid line is the mean profile and the 1$\sigma$ upper and lower
limits for 100 simulations of an isothermal spherical cluster having
$\sigma=1460$ km s$^{-1}$. The input model is shown by the dotted line. The
lower panel is the radial density profile from Eqn. \protect\ref{denseqn} using
$\kappa$ derived from fits to the simulations and incorporating a radially
varying $\Sigma_{crit}$ as shown in Fig. \protect\ref{sigc}. A recovery factor
of $C = 3.0$ (see text) has been applied.
\label{profilef}}
\end{figure}

Because of the smoothing kernel, plus biases introduced by edge effects in the
images, Eqn \ref{kseqn} is mainly useful for determining the 2-d shapes of mass
distributions. A less biased way of obtaining mass estimates as well as
azimuthally averaged density profiles is:

\begin{equation} \label{denseqn}
\overline\kappa(r \le r_i) - \overline\kappa(r_i \le r \le r_o) =
{r_o^2 \over N_{io}}\sum_{r_i \le r \le
r_o}{[1-\kappa(\vec{r})][1-\sqrt{1-D_i(\vec{r})^2}]
\over D_i(\vec{r})(\Delta{x_i}^2 + \Delta{y_i}^2)}
\end{equation}

\noindent
where $N_{io}$ is the number of galaxies between $r_i$ and $r_o$.
This is similar to the form employed by Fahlman et al. (1994) but is valid when
$\kappa$ is not vanishingly small. Since $\kappa$ appears on the right hand
side of the equation, an iterative approach must be used to obtain the density
profile.  Radial mass profiles for RXJ1347.5-1145 are shown in
Fig. \ref{profilef} and are discussed further in \S \ref{profile}.

It should be mentioned that galaxy distortion is insensitive to flat sheets of
mass. Consequently, all mass measurements described in this paper are uncertain
by an unknown additive constant. If there is a substantial flat component to
the mass distribution our mass estimates will be lower limits. This is
discussed further in \S \ref{profile}.

\subsection{Monte-Carlo Simulations} \label{calibration}

Before proceeding to a discussion of the mass reconstruction for the cluster we
describe Monte-Carlo simulations of the data. These simulations are useful for
identifying sources of systematic error which arise from seeing, measurement
error and deviations from the weak lensing approximations, and for calibrating
the data.

The simulations are discussed in detail in Fischer et al. (1996) and we briefly
summarize them here. The simulations consist of artificial galaxies distributed
in seven redshift shells (z = 0.0 - 0.45, 0.45 - 0.6, 0.6 - 0.7, 0.7 - 0.8, 0.8
- 1., 1.0 - 1.4, 1.4 - 7.0). Galaxy images are generated for each shell based
on the quiescent models of McLeod \& Rieke (1995). Stars are also added to the
simulated images based on the observed star counts. The size-absolute magnitude
relationship for the simulated galaxies is adjusted to yield the same recovery
factor as the HDF simulations and the simulated galaxies have a similar
apparent ellipticity distribution to the observed galaxies. The galaxies in
each shell are distorted with various spherically symmetric mass distributions
located at $z=0.451$ under the assumption that all galaxies within a redshift
shell lie at the number weighted mean redshift of that shell. Convolution with
the PSF and the addition of noise are followed by identical processing to the
real data. The results of the simulations will be discussed in the next
section.

\subsection{2-d Mass Maps} \label{2d}

The 2-d, KS mass map for RXJ1347.5-1145 is shown in Fig. \ref{massmap}
superposed on the R-band image of the field.  This reconstruction used 2735
galaxies with $23.0 \le$ B$_J \le 25.0$. The central peak has a signal-to-noise
ratio of around 11 with $s = 43\arcsec$, and is consistent with the position of
the CDG. The noise is calculated from maps derived from 100 simulations
described above.

To check for biases introduced by residual anisotropies in the PSF, a similar
map was produced using 193 stars detected in the field. The maximum in that map
is 9\% of the peak value in the cluster mass map (equal to the 1.0$\sigma$
noise level of the mass map) and the minumum is -6\%.  We conclude that
residual PSF anisotropy is not significantly affecting the 2-d massmap. For
comparison a map made from the same stars on the uncorrected image has a
maximum of 17\% of the mass map peak and minimum of -12\%, and had large scale
correlations not present in the corrected map.

\begin{figure}
\plotone{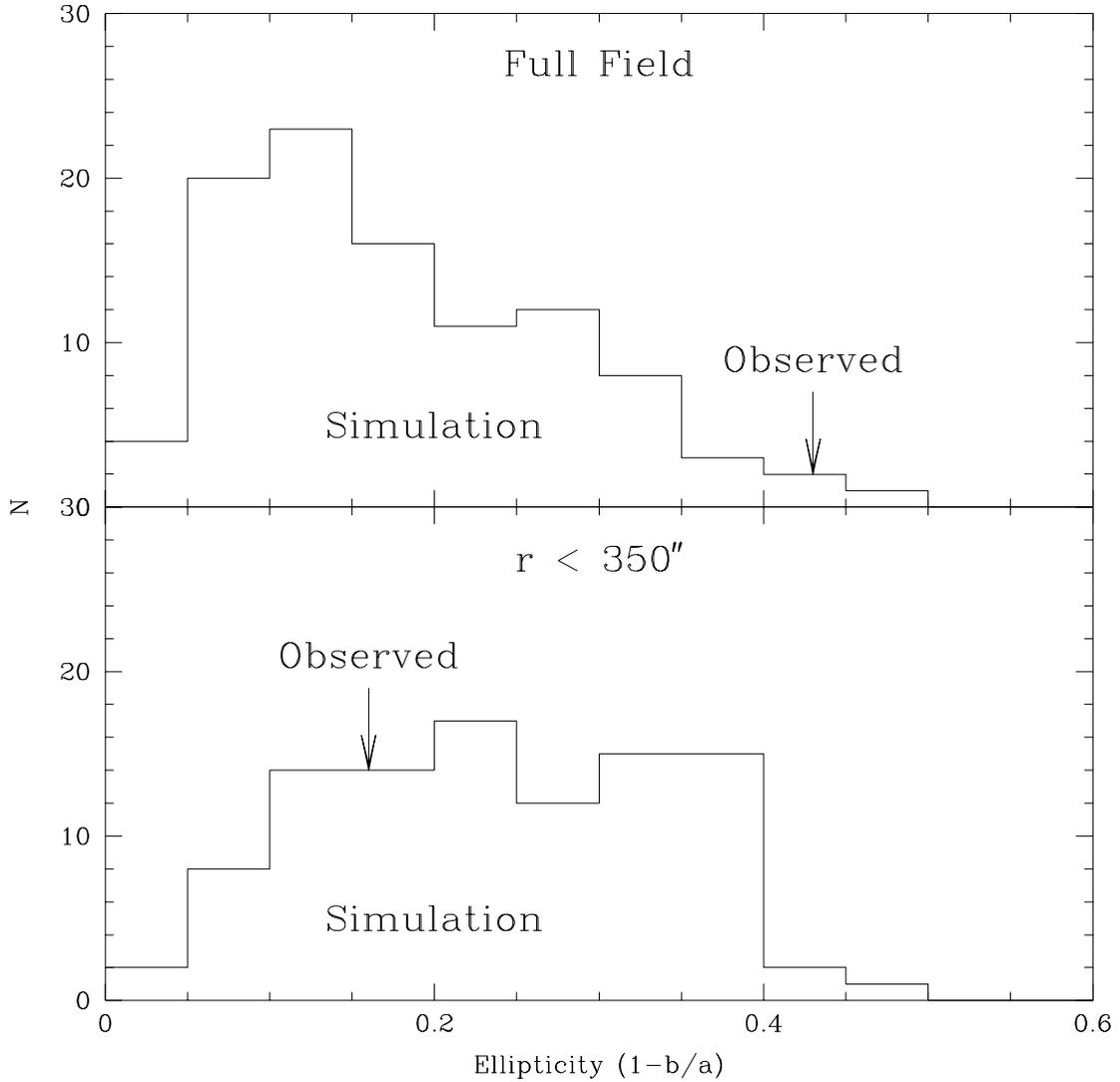}
\caption{Histogram of ellipticity values for KS mass maps
of 100 simulations of spherical isothermal lens ($s = 43\arcsec$). The
ellipticities are derived from quadrupole measurements of the positive portions
of the mass maps. The upper panel shows the values for the full CCD field, the
observed value for the RXJ1345.5-1145 field is indicated by the arrow. The
cluster is more elongated than 99 out of 100 simulations. The lower panel shows
the values for the region within 350\arcsec\ of the CDG; the cluster falls
within the range of the simulations in this region.
\label{quad}}
\end{figure}

The detectability of substructure in the mass map is limited by the surface
density of background galaxies which is approximately 13 arcmin$^{-2}$ in this
case. The surface mass density contours appear to be fairly circular in the
cluster central region and there are two 3$\sigma$ features in the southwest
corner separated by about 20\arcsec = 70 $h^{-1}$ kpc.  One worries about
over-interpreting the KS mass map since the technique is known to produce
spurious features (Schneider 1995). In order to test the significance of
deviations from circularity in the KS mass reconstruction we have adopted the
approach of Fischer et al.  (1996).  Ellipticity is computed from the
quadrupole moments of the portions of the map with $\kappa >0$.  The upper
panel of Fig. \ref{quad} shows this ellipticity value along with the
distribution of ellipticities measured in an identical manner from the
Monte-Carlo simulations, each of which contain spherical lenses. The RXJ
1347.5-1145 mass map is more elongated than 99 out of 100 simulations
(Fig. \ref{quad}). When we restrict the quadrupole moment calculation to $r \le
350\arcsec$ we find the cluster ellipticity falls within the range seen in the
simulations. Therefore, the large ellipticity measured in the full field map is
due to the subcluster in the southwest corner. We conclude that the subcluster
is significant and unlikely to be an artifact of the KS image
reconstruction. The subclumps are 385\arcsec\ and 500\arcsec\ from the cluster
center (1.3 and 1.7 $h^{-1}$ Mpc, respectively).  Of course, the redshifts of
the subclumps are unknown, and since the mass map is sensitive to mass over a
large range of redshifts these subclumps may not be associated with the main
cluster.

Fig. \ref{galdens} is a contour plot of galaxy number density for galaxies in
the range B$_J \ge 20.45$ and B$_J$ -- R $> 1.6$. The center of the galaxy
density map is consistent with the mass center and the position of the CDG. The
galaxy map does not show evidence for subclustering in the southwest in this
color range, however, both mass and galaxy maps do show evidence for extensions
leading from the cluster center to the southeast.

\begin{figure}
\caption{Number density contours of galaxies having $B_J-R
\ge 1.6$ superposed on the R band image of RXJ1347.5-1145. The contours are
spaced by 1$\sigma$ (2.4 arcmin$^{-2}$) about the mean density (7.2
arcmin$^{-2}$). Orientation and size are as in Fig. \protect\ref{massmap}
\label{galdens}}
\end{figure}

\subsection{Mass Density Profile} \label{profile}

In the upper panel of Fig. \ref{profilef} we show the azimuthally averaged
surface mass density profile centered on the CDG as derived from
Eqn. \ref{denseqn} using 1970 galaxies having 23.0 $\le $B$_J \le 25.0$. We
have assumed that $\kappa = 0$ on the right hand side and that the critical
density is constant ($\Sigma_{crit} = 9700 h$ M$_\odot$ pc$^{-2}$). Also shown
is the mean measured mass profile for 100 simulations of a singular isothermal
lens with $\sigma = 1460$ km s$^{-1}$ (see \S \ref{calibration}) which matches
the radial profile for RXJ1347.5-1145 quite well. The measured simulated
profile derived in this manner is much steeper than the input profile. However,
if we now plug in the known values for $\kappa$ (based on the input density
profile and $\Sigma_{crit}$ from Fig. \ref{sigc}) we get much better agreement
between the input and derived density profiles for the simulationss (bottom of
Fig. \ref{profilef}). The innermost point is underestimated but this is the
region where the magnification is the highest and the lensing effect is the
strongest so we expect the largest systematic errors. Correcting the
RXJ1347.5-1145 data in a similar way yields the points shown in the lower panel
of Fig. \ref{profilef}. Therefore, the cluster profile appears to be consistent
with a singular isothermal sphere having velocity dispersion $\sigma = 1460 \pm
150$ km s$^{-1}$.

Note that the densities in Fig \ref{profilef} are plotted as density contrasts;
the mean density within a radius minus the mean density in a control
annulus. The latter is estimated by extrapolating the density profile beyond
the measured region so it is worth mentioning the relative values of the
two. Based on the above mentioned singular isothermal model the control annuli
densities vary from 8\% of the total density for the innermost point to 50\%
for the outermost point.

The weak lensing signal will be diluted by cluster galaxies, and this dilution
will be larger in the central regions. This will result in a systematic
underestimate in the cluster mass, and will also flatten the inferred density
profile. In order to estimate this effect we have looked at galaxies in the
range 23.0 $\le $B$_J \le 25.0$ (the range used in the mass reconstruction) and
compared the numbers of red galaxies having B$_J$ -- R $\ge 1.6$ (see below)
with the total number. We have assumed that the number density of the red
galaxies at the edge of the field is dominated by field galaxies and have
subtracted the mean density there (1.75 arcmin$^{-1}$) from the counts. We
estimate that the sample of galaxies used to produce the innermost point in the
radial density plot contains about 9\% cluster galaxies declining to about 5\%
at large radius. These galaxies should be randomly aligned and therefore will
have zero shear. If these galaxies have been included in the sum in
Eqn. \ref{denseqn} then $N_{io}$ will be overestimated and the density will be
underestimated. Applying a correction for these cluster galaxies slightly
steepens the density profile (although it is still consistent with the
isothermal density profile) and increases the velocity dispersion to $\sigma =
1500 \pm 160$ km s$^{-1}$.

\begin{figure}
\plotone{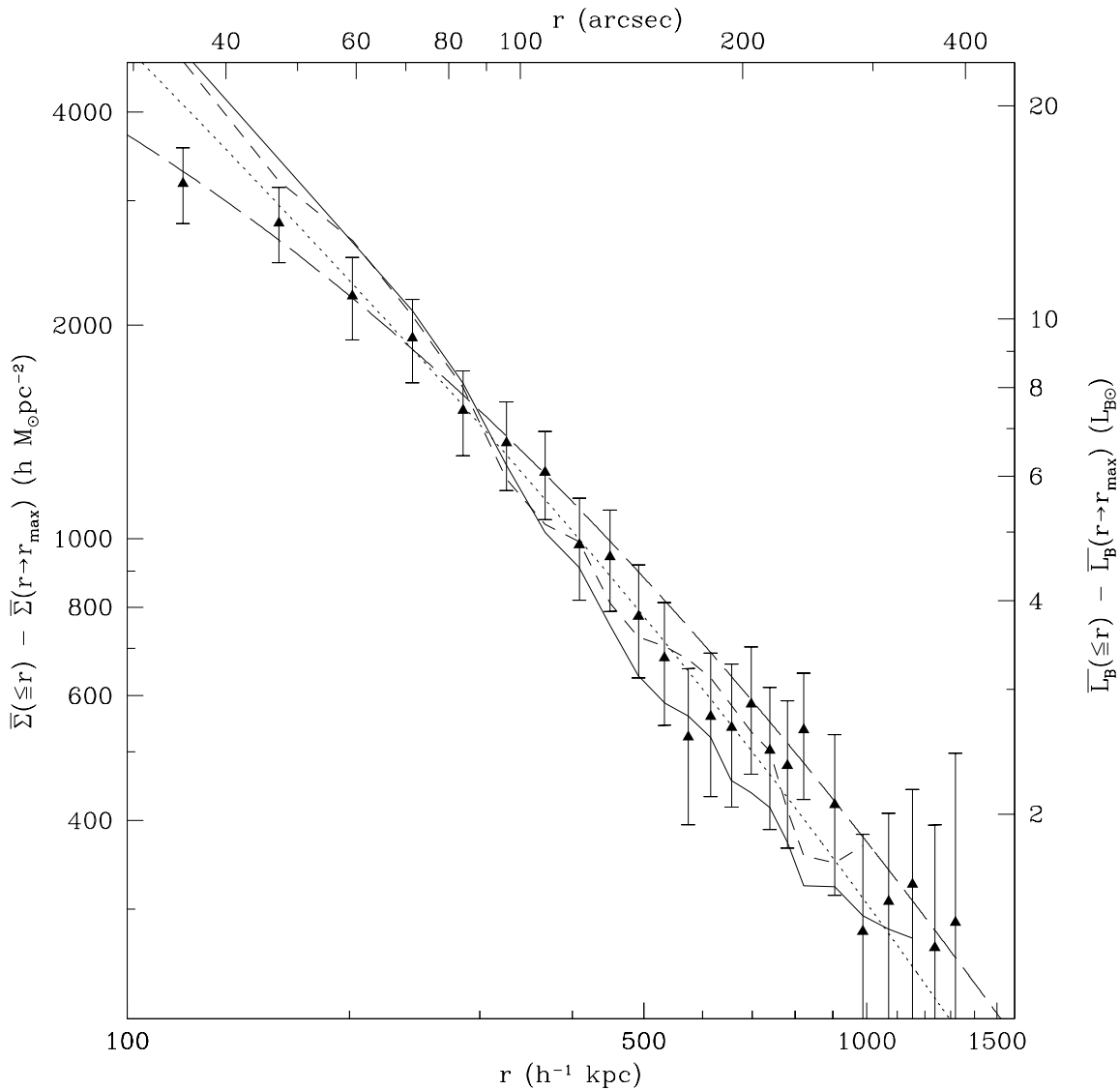}
\caption{Plot of projected total cluster mass density and
projected rest-frame B-band luminosity in galaxies. The luminosity density is
shown for B$_J \ge 20.45$ and $B_J$ -- R $\ge 1.6$ (solid) and $B_J$ -- R $\ge
0$ (short-dashed). The mass density is shown as points.  Both densities are
plotted as density contrasts. The dotted line is a singular isothermal model
with $\sigma = 1500$ km s$^{-1}$ and the long-dashed line is the profile of
Navarro et al (1995) with $R_{200} = 1.2 h^{-1}$ Mpc. The mass and luminosity
densities are consistent over the entire radial range plotted, with rest-frame
M/L$_B = 200 \pm 50 h$ M$_\odot$/L$_{R\odot}$. No significant difference is
seen between the two different luminosity profiles indicating that the cluster
light is dominated by galaxies with observed B$_J - $R $\ge$ 1.6.
\label{surf}}
\end{figure}

In Fig. \ref{surf} we show the corrected surface mass density along with an
isothermal profile having $\sigma=1500$ km s$^{-1}$. Also plotted is the
``universal'' density profile of Navarro et al. (1995) which in 3-d is given
by:

\begin{equation}
{\rho(R) \over \bar{\rho}} = {1500r^3_{200} \over R(5R+R_{200})},
\end{equation}

\noindent
where $\bar{\rho} = 2.78 \times 10^{-7}(1+z)^3 h^2$ M$_\odot$ pc$^{-3}$ is the
critical density of the universe and $R_{200}$ is the radius for which the mean
interior overdensity is 200. The curve shown in Fig. \ref{surf} for the surface
density has $R_{200} = 1.2 h^{-1}$ Mpc and is consistent with the data. In
order to distinguish between these two density laws one would need data
extending out to larger radius than we have here as the Navarro et al. surface
density profile becomes steeper than an isothermal at large radius.



\subsection{Mass-to-Light Ratio}

Also in Fig. \ref{surf} we show the rest-frame B-band light contained in
galaxies with R $\ge 18.0$ for both B$_J$ -- R $>$ 1.6 and B$_J$ -- R $>$
0.0. There is little difference in the two light profiles, indicating that the
cluster light is dominated by galaxies with observed B$_J - $R $\ge$ 1.6
(actually this only indicates that the radially varying component of the
cluster light is dominated by red galaxies). We have only included light in
galaxies out to an isophotal level of R = 29.8 mag per square arcsec, any
component more extended than this (for example a diffuse cluster component) is
not included in the plotted luminosity profile or the mass-to-light ratio
calculations described below.

The rest-frame B-band light is derived from the observed R-band light,
requiring only a small correction, since, for z = 0.45, the central wavelength
of the B bandpass (4490 \AA) redshifts almost exactly to the center of R (6510
\AA). There are two correction terms, the first due to redshift stretching of
the bandpass and the second due to the relative calibration of the two
bandpasses (based on the flux of A0 star). The absolute rest-frame B magnitude
is given by:

\begin{equation}
M_B  =  R - 5\log(D_l) +5 - 2.5\log(1+z) - 2.5\log[f_R(A0)/f_B(A0)], \\
\end{equation}

\noindent where $D_l$ is the luminosity distance in $h^{-1}$ parsecs, and
$f_R(A0)/f_B(A0)$ is the relative flux of an A0 star in the two bandpasses.

The luminosity density has been plotted as a density contrast similarly to the
mass density.  The mass and light profiles are consistent with one another for
the entire radial range shown, with marginal evidence that the light profile is
slightly steeper. They are also both consistent with a singular isothermal
model. For $r<35\arcsec$ the luminosity density is steeper than isothermal (Not
shown).

The implied rest-frame mass-to-light ratio is M/L$_B = 200 \pm 50 h$
M$_\odot$/L$_{B\odot}$ (M$_{B\odot} = 5.48$). If we apply the estimated
k-correction for early-type galaxies directly to the measured R-band light
[$k_{corr} = 0.625$ mag (Poggianti 1996)], we get M/L$_R = 150 \pm 40 h$
M$_\odot$/L$_{R\odot}$ (M$_{R\odot} = 4.32$).  According to Poggianti (1996)
the estimated evolutionary correction for early type galaxies is $B = -0.62$
mag for a $z=0.451$ galaxy, which agrees fairly well with the measured value
for cluster ellipticals from Schade et al. (1997) of about $0.47 \pm
0.15$. Adopting these values and correcting to $z=0$ we get M/L$_B = 310 - 350
\pm 90 h$ M$_\odot$/L$_{B\odot}$.  Hughes (1989) found M/L$_B = 280 - 360 h$
M$_\odot$/L$_{B\odot}$ for the Coma cluster based on X-ray measurements and the
assumption that mass traces light [in agreement with dynamical mass estimate of
Colless \& Dunn (1996)]. Therefore, RXJ1347.5-1145, despite being more than a
factor of two more massive than Coma, appears to have a similar fraction of its
mass in stars.




\subsection{Arc Candidates}

With an estimate of the mass density profile we can now predict the redshift of
the two arc candidates, discovered by Schindler et al. (1995), under the
assumption that they are strongly lensed background galaxies located at the
Einstein radius. The arc candidates are located at projected distances of
34.4\arcsec\ and 36.0\arcsec\ from the CDG. The former has B$_J = 23.0$ and
B$_J$ -- R = 1.1 while the latter has B$_J = 21.9$ and B$_J$ -- R = 0.4 and has
a lower surface brightness; both are significantly bluer than the CDG. We
estimate that, in order of distance from the cluster center, the arc redshifts
are $z=1.4^{+1.40}_{-0.35}$ and $z=1.6^{+2.00}_{-0.50}$, respectively. It
should be mentioned that aside from being tangentially aligned to the
center of the cluster, neither of these galaxies appears to be particularly
arc-like (they are not highly elongated and no bending is apparent in our
images) so the possibility remains that they are foreground field galaxies.


\subsection{Comparison With X-ray Mass}

There is a mass estimate for RXJ 1347.5-1145 based on X-ray data from both the
ROSAT and ASCA satellities (Schindler et al 1996). This study used ROSAT HRI
data to determine the shape of the radial profile and ASCA GIS data to measure
the temperature. The cluster mass was estimated using the standard
$\beta$-model technique (Cavaliere \& Fusco-Femiano 1976), assuming hydrostatic
equilibrium, spherical symmetry and isothermality, and was found to be M$(R<1
h^{-1}$ Mpc) = $5.8 \times 10^{14} h^{-1}$ M$_\odot$ with a 15-20\% quoted
uncertainty. Our value from gravitational lensing is M$(R<1 h^{-1}$ Mpc) = $1.1
\pm 0.30 \times 10^{15} h^{-1}$ M$_\odot$, a factor of 1.8 higher than the
X-ray mass estimate.

This mass difference is similar to what has been seen in comparisons between
strong lensing and X-ray mass estimates at small projected radius for Abell
1689 and Abell 2218 (Miralda-Escude \& Babul 1995), and between the
weak-lensing and X-ray mass estimates of 0957+561 (Fischer et
al. 1996). However, the weak lensing and X-ray masses for Abell 2218 are
consistent out to $0.4 h^{-1}$ Mpc (Squires et al. 1996a) and the same is true
for Abell 2163 after the X-ray mass was used to estimate the mass density in
the control annuli (Squires et al. 1996c).

The lensing mass estimate will be an overestimate if we have underestimated the
mean redshift of the background galaxies. If we assume that all the background
galaxies are at $z=3.0$, $\Sigma_{crit}$, and hence the inferred cluster mass,
drops by about a factor of two (actually less than this since for part of our
radial range we are assuming a lower $\Sigma_{crit}$, see Fig. \ref{sigc}). 
However, based on recent Keck redshift surveys, this is not possible. For
example, Cowie et al. (1996) find a median redshift of around $z \approx 0.8$
at B = 24.0, close to our adopted value for the range B$_J = 23.0 - 25.0$.

If one accepts the X-ray mass then one is led to the conclusion that this
cluster is somewhat anomolous in several respects. The X-ray mass yields M/L$_B
= 110 h$ M$_\odot$/L$_{B\odot}$, lower than seen in any other cluster. The
value for M/L$_X$(bol) [L$_X$(bol) $= 2\times 10^{46}$ erg s$^{-1}$] is much
lower than is typically seen in other clusters (Sarazin 1988, Fig. 28), and the
temperature (9.3 keV) is 35\% lower than would have been expected for a cluster
having the X-ray bolometric luminosity of RXJ 1347.5-1145 (David et al
1993). If the cluster temperature is underestimated one possible explanation is
that it has been biased by an unresolved cooling flow. Schindler et al (1996)
cite evidence for a large cooling flow, but due to the resolution limits of the
ASCA data are not able to measure a central temperature decrement, and do not
attempt to correct their temperature estimate. Finally, with the X-ray mass
estimate, the gas mass fraction is very high, 34\% within 1 Mpc and 52\% within
3 Mpc ($h = 0.5$). For the lensing mass estimate, the gas mass fraction is
reduced by a factor of 1.8, and falls within the range seen for other rich
clusters within similar radii (White \& Fabian 1995, Buote \& Canizares 1996).

Currently there is considerable debate regarding the reliability of X-ray mass
estimates. X-ray masses require the assumption of hydrostatic equilibrium, but
cluster-cluster mergers can cause large deviations from this state.  This in
turn can lead to mass errors of $>100\%$ right after a merger and 50\% for up
to 2 Gyrs afterwards (Roettiger et al. 1996). Since predictions for typical
cluster merger rates are approximately 1 every 2-4 Gyr (Edge et al.  1991)
these sorts of X-ray mass errors are not unexpected for any given cluster
measurement. Simulations following the formation of clusters in different
cosmologies reveal a 1$\sigma$ scatter of 30\% in cluster mass measurements
based on the $\beta$-model (Evrard et al. 1996), indicating the uncertainties
quoted by Schindler et al. (1996) are probably underestimated.  Furthermore the
presence of substructure generally results in X-ray masses which are
underestimated if a spherical cluster is assumed. Finally, the difference in
mass estimates could be reduced if the cluster is elongated along the
line-of-sight, resulting in an underestimated X-ray mass. 

\section{Conclusion}

In this paper we study the mass distribution of the $z=0.451$ X-ray cluster
RXJ1347.5-1145 out to projected radii of 1.4 $h^{-1}$ Mpc by measuring the
gravitationally-induced distortions of background galaxies. We detect a shear
signal in the background galaxies in the radial range $35\arcsec\ \le r \le
400\arcsec$ significant at the 7.5$\sigma$ level. The resultant mass map
exhibits an 11$\sigma$ peak centered on the dominant cluster galaxy. There is
evidence for subclustering in mass at 1.5 - 2.0 $h^{-1}$ Mpc from the cluster
center. However, since the weak lensing mass estimates are sensitive to mass
over a large redshift range it is not certain that the subclusters are
associated with the main cluster. No corresponding excess is detectable in the
galaxy counts.

The azimuthally averaged mass and light profiles follow one another and are
consistent with a singular isothermal model for the radial range $120 \le r \le
1360$ ($h^{-1}$ kpc). They are also consistent with the ``universal'' density
profile of Navarro et al. (1995).  The implied velocity dispersion (assuming
isotropic distribution function) is $\sigma = 1500 \pm 160$ km s$^{-1}$. The
lensing mass estimate is almost two times higher than the recent X-ray mass
estimate of Schindler et al. (1996). This large difference might be
attributable to some combination of a recent merger, substructure, and
elongation along the line-of-sight. The weak lensing mass estimate yields a
rest-frame mass-to-light ratio of M/L$_B = 200 \pm 50 h$
M$_\odot$/L$_{B\odot}$. After estimated evolution correction this corresponds
to M/L$_B = 310 - 350 \pm 90 h$ M$_\odot$/L$_{B\odot}$ at $z=0$, similar to what is
seen in the Coma cluster, despite being over two times more massive.

Similar mass traces light behavior has been seen in other weak lensing studies
(Tyson \& Fischer 1995, Squires et al. 1996a, 1996b, 1996c) and in a radial
velocity study of a sample of 16 clusters Carlberg et al (1996). These results
argue against a significant velocity bias between the galaxies and the dark
matter particles.  Many simulations that follow the evolution of hierarchical
galaxy formation in clusters show some degree of velocity bias (i.e. Carlberg
1994, Frenk et al. 1996). A direct comparison between the simulations and the
cluster results is not straighforward due to the difficulty in identifying
galaxies in the simulations. A futher difficulty is that the simulations show
the greatest velocity bias in the inner regions where the X-ray masses are the
most in doubt and the lensing masses are incomplete. Therefore, it is probably
fair to say that most simulations are not grossly inconsistent with the
mass-traces-light result but any simulations which predict a large bias over
the relevant radial ranges are suspect.

\acknowledgments

Support for this work was provided by NASA through grant \# HF-01069.01-94A
from the Space Telescope Science Institute, which is operated by the
Association of Universities for Research in Astronomy Inc., under NASA contract
NAS5-26555. Thanks to Gary Bernstein for assistance in devising the
circularizing kernel and for reading an earlier draft of this paper.  Thanks to
Joe Mohr for interesting discussions/arguments. We also thank the referee, Gus
Evrard for many helpful comments which improved the paper.

{}



\end{document}